\documentstyle[12pt,epsf,subfigure,color,ulem]{article}

\newcommand{\cd}{\makebox[0.08cm]{$\cdot$}}

\title{Critical stability of few-body systems}
{\author{V.A.~Karmanov$^a$ and J.~Carbonell$^b$\\
{\small \em  $^a$Lebedev Physical Institute, Moscow, Russia}
\\
{\small \em $^b$Institut de Physique Nucleaire, Orsay, France}
}}
\begin{document}
\maketitle
\bibliographystyle{unsrt}

\begin{abstract}
When a  two-body system is bound by a zero-range interaction,
the corresponding three-body system -- considered in a non-relativistic framework --
collapses, that is its binding energy is unbounded from below. 
In a paper by J.V.~Lindesay 
and H.P.~Noyes \cite{Noyes} it was shown that the relativistic effects result in an effective repulsion 
in such a way that three-body binding energy remains also
finite, thus preventing the three-body system from collapse. Later, this property was confirmed in other works based on different versions of relativistic 
approaches. However, the three-body system exists only for a limited range of two-body binding energy values. 
For stronger two-body interaction, the relativistic three-body system still collapses. 

A similar phenomenon was found in a two-body systems themselves: a two-fermion system with one-boson exchange interaction in a state with zero angular momentum $J=0$ exists if 
the coupling constant does not exceed some critical value but it also collapses for larger coupling constant. 
For a $J=1$ state, it collapses for any coupling constant value. These properties are called "critical stability". 
This contribution aims to be a brief review of this field  pioneered by H.P. Noyes.
\end{abstract}

%%%%%%%%%%%%%%%%%%%%%%%%%%%%%%%%%%%%%%%%%%%%%%%%%%%
\section{Introduction}\label{intr}

The radius of nuclear forces -- the interaction between  protons and neutrons  -- is sensibly smaller than the size of nuclei themselves. 
Since the wave function at large distances $r$ behaves as $\sim \exp(-|E_b| r)$, the latter is determined by the nuclear binding energy $E_b$. 
The binding energy, in its turn, is a cancellation of a large (negative) potential energy and large (positive) kinetic
energy. Therefore $E_b$ is much smaller than each of these energies and the nuclear radius
$r\sim 1/|E_b|$ can be larger than the radius of the nuclear forces. To understand qualitatively some nuclear properties, 
one can consider the "zero-range interaction limit". 
To this aim, we approximate the nuclear interaction $V$ by a potential well:
$$
V(r)=\left\{
\begin{array}{ll}
-U_0,& \mbox{if $r<a$}
\\
\phantom{-}0, & \mbox{if $r>a$}
\end{array}
\right.
$$
As it is well known from standard quantum mechanics textbooks (see e.g. \cite{ll}), 
a bound state exists  if some relation between the potential depth $U_0$ and its range $a$ is fulfilled, that is if
\[ U_0> {\pi^2 \hbar^2\over 8ma^2} \]
$\hbar$ being the Plank constant and $m$  the mass of the particle. 
If we let $a$ tend to zero and  $U_0$ to infinity, keeping constant the product $U_0a^2$, we will get in this limit an infinitely deep zero range potential well, in which a two-body bound state exists. 

\begin{figure}[!ht]
\begin{center}
\mbox{\epsfxsize=8.cm\epsffile{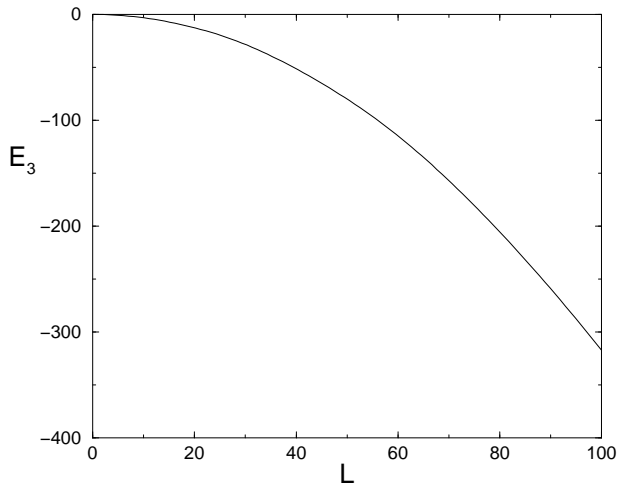}}
\caption{Three-body binding energy (in the units of mass $m$), for the zero-range two-body  interaction and finite two-body binding energy, as a function of momentum cutoff $L$ in the  Faddeev equation.\label{fig3b}}
\end{center}
\end{figure}

The zero-range two-body interaction provides an important limiting case which
qualitatively reflects characteristic properties of nuclear \cite{Nuclear} and atomic \cite{Atomic} few-body systems. 
It turned out, however, that when using non-relativistic dynamics, it generates the Thomas collapse \cite{thomas} of the three-body system.
The latter means that the three-body binding energy tends to $-\infty$, when the interaction radius tends
to zero keeping constant the product $U_0a^2$ and consequently the two-body binding energy.
As an illustration, we have solved numerically the three-body Fadeev equation in momentum space  
with the two-body amplitude for zero-range interaction as input. The corresponding three-body
binding energy is kept finite by introducing a momentum cutoff $L$. 
The result for the three-body binding energy (in units of the particle mass $m$) is shown in fig.~\ref{fig3b}. 
We see that when cutoff $L$ is removed ($L$ tends to infinity), the three-body binding energy $|E_3|$ increases monotonously without any limit. 
This is just the manifestation of the Thomas collapse. Several ways to regularize this interaction have been proposed in the literature \cite{FRED,FJ}.

It should be emphasized that the Thomas collapse was found in the non-relativistic framework, which should be applied only when the binding energy is 
much smaller than the particle mass. We see that the results displayed in fig.~\ref{fig3b}  do not correspond to this situation:
the module of the binding energy $|E_3|$ becomes much  larger that particle mass. For example, for the cutoff $L\approx 50\,m$ the binding energy is $E_3\approx - 100\,m$. 
This is far beyond the domain where the non-relativistic treatment is valid. The answer to the question: "what happens with the three-body system in the limit of two-body zero-range interaction"
should be obtained in a relativistic framework only.

This answer was first found in the paper by J.V.~Lindesay and H.P.~Noyes \cite{Noyes} in the so called ''minimal relativistic model".
It was shown that the relativistic effects result in an effective repulsion and can thus prevent the three-body system from collapse: the three-body binding energy remains also finite. 

Later, this property was confirmed in other works based on different versions of relativistic 
approaches. In particular, two-body calculations showed
that in the scalar case, relativistic effects were indeed stronglly repulsive \cite{MC_PLB_00}. 
However, it was found \cite{3bosons} that this stabilization]had some restrictions: the three-body system exists only in a limited range of two-body binding energy. 
For stronger two-body interaction, the mass squared of three-body system $M_3^2$ though remaining  finite, crosses zero and becomes negative. 
This means  that the relativistic three-body system does not longer exists.  
Then a similar phenomenon was also found  in the two-body systems: the two-fermion systems with one-boson exchange interaction also 
collapses if the coupling constant exceeds some critical value. 
These properties are called "critical stability" and they are forming now an interesting field of research.
In what follows we will give a brief review of this developing field  pioneered by H.P. Noyes.

%%%%%%%%%%%%%%%
\section{Relativistic three-body system with zero-range interaction}

In  paper \cite{Noyes},  relativistic three-body calculations with zero-range
interaction have been performed in a  minimal relativistic model.  Later, a much more general and sophisticated 
approach to the relativistic few-body systems -- Light-Front Dynamics -- was developed (see for review 
\cite{cdkm,BPP_PR_98}). In the framework of this relativistic approach the problem of three equal mass ($m$) bosons interacting via zero-range forces was reconsidered in the \cite{3bosons}.

The relativistic three-body equation is derived in Section \ref{equat}. In
Section \ref{results}, their solutions are presented and some concluding remarks are given in Section \ref{concl}.

%%%%%%%%%%%%%%%%%%%%%%%%%%%%%%%%%%%%%%%%%
\subsection{Equation}\label{equat}

Our starting point is the explicitly covariant
formulation of the Light-Front Dynamics  \cite{cdkm}. In non-relativistic approach the wave function $\psi(\vec{r},t)$ 
is a probability amplitude defined at a given time  $t$, say at $t=0$. In four-dimensional Minkowski space 
one can define the wave function on any space-like plane to preserve the causality, or more generally  on any space-like surface. 
The orientation of this plane is defined by a four-vector $\lambda=(\lambda_0,\vec{\lambda})$  orthogonal to this plane. 
We can change its orientation moving $\lambda$ within the light cone in such a way that the plane where the wave function is defined remains space-like.
Its limiting value is reached when $\lambda$ lies on the light-cone surface. 
Then such the four-vector is denoted by  $\omega=(\omega_0,\vec{\omega})$ and has the property $\omega^2=\omega_0^2-\vec{\omega}^2=0$.

The corresponding plane  is given by the equation $\omega\cd x=0$. In the particular case $\omega=(1,0,0,-1)$ it turns into $t+z=0$, seting hereafter $c=1$. 
This equation coincides with the light-front equation and therefore the plane $t+z=0$ is called the light-front plane. 
The dynamics determining the  evolution of the wave function from one light-front plane to another one is the light-front dynamics. 
This approach was proposed by Dirac \cite{Dirac} and it has many advantages. 
Later,  its explicitly covariant version was developed, when the light-front plane is defined by the covariant equation $\omega\cd x=0$ and no any particular axes like $t$ or $z$ 
is selected \cite{cdkm}. We will just use the light-front dynamics as a relativistic approach.

\begin{figure}[!ht]
\begin{center}
\mbox{\epsfxsize=15.cm\epsffile{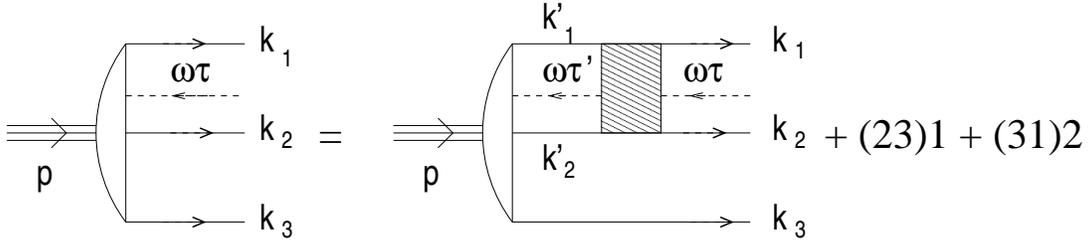}}
\caption{Three-body equation for the vertex function $\Gamma$.\label{fig1}}
\end{center}
\end{figure}

The three-body equation is represented graphically in figure \ref{fig1}.
It concerns the vertex function $\Gamma$,
related to the wave function $\psi$ in the standard way:
$$
\psi(k_1,k_2,k_3,p,\omega\tau)=\frac{\Gamma(k_1,k_2,k_3,p,\omega\tau)}
{{\cal M}^2-M^2_3},\quad
{\cal M}^2=(k_1+k_2+k_3)^2=(p+\omega\tau)^2.
$$

All four-momenta are on the corresponding mass shells ($k_i^2=m^2$,
$p^2=M_3^2,$ $(\omega\tau)^2=0$) and satisfy the conservation law
$k_1+k_2+k_3=p+\omega\tau$ involving  $\omega\tau$. The four-momenta
$\omega\tau$ and $\omega\tau'$ are drawn in figure \ref{fig1} by dash lines.
The off-energy shell character of the wave function is ensured by non-zero value of the scalar
variable $\tau$.  In the standard approach \cite{BPP_PR_98}, the  minus-components of the momenta
are not conserved and the only non-zero component of $\omega$ is
$\omega_-=\omega_0-\omega_z=2$. Variable $2\tau$ is just the non-zero
difference of non-conserved components $2\tau=k_{1-}+k_{2-}+k_{3-}-p_-$.

Applying to figure \ref{fig1} the covariant light-front graph techniques \cite{cdkm}, we find the equation:
\begin{eqnarray}\label{eq3}
\Gamma(k_1,k_2,k_3,p,\omega\tau) 
&=& \frac{\lambda}{(2\pi)^3} \int \frac{d\tau'}{\tau'}\,
\frac{d^3k'_1}{2\varepsilon_{k'_1}}\,
\frac{d^3k'_2}{2\varepsilon_{k'_2}}
 \;\; \Gamma(k'_1,k'_2,k_3,p,\omega\tau') \nonumber\\
&\times&
\delta^{(4)}(k'_1+k'_2-\omega\tau'-k_1-k_2+\omega\tau)
\;+\; (23)1+(31)2,
\end{eqnarray}
where  $\varepsilon_{k}=\sqrt{m^2+\vec{k}^2}$.
For the zero-range forces we are interested in, the interaction kernel appears as a constant $\lambda$.
In (\ref{eq3}) the contribution of interacting pair (12)  is explicitly written while
the contributions of the remaining pairs are simply denoted by $(23)1+(31)2$.

Equation (\ref{eq3}) can be  rewritten in variables $\vec{R}_{i\perp},x_i,$ ($i=1,2,3$),
where $\vec{R}_{i\perp}$ is the spatial component of the four-vector
$R_i=k_i-x_ip$ orthogonal to $\vec{\omega}$ and $x_i={\omega\cd k_i\over\omega\cd p}$ \cite{cdkm}.
For this aim we insert in r.h.-side of (\ref{eq3}) the unity integral
$$
1=\int
2(\omega\cd k'_3)\delta^{(4)}(k'_3-k_3-\omega\tau_3)d\tau_3\,
\frac{d^3k'_3}{2\varepsilon_{k'_3}}
$$
and recover the usual three-body space volume which, expressed in the variables $(\vec{R}_{i\perp},x_i)$, reads
\[
\int
\delta^{(4)}(\sum_{i=1}^3 k'_i-p-\omega\tau')
\prod_{i=1}^3{d^3k'_i\over2\varepsilon_{k'_i}} 2(\omega\cd p) d\tau'
=\int\delta^{(2)}(\sum_{i=1}^3\vec{R'}_{\perp i})
\delta(\sum_{i=1}^3 x'_i-1)2\prod_{i=1}^3{d^2R'_{\perp i}dx'_i\over2x'_i}.\]

The Faddeev amplitudes $\Gamma_{ij}$ are introduced in the standard way:
$$\Gamma(1,2,3)=\Gamma_{12}(1,2,3)+\Gamma_{23}(1,2,3)+\Gamma_{31}(1,2,3), $$
and equation (\ref{eq3}) is equivalent to a system of three coupled equations for these components.
With the symmetry relations $\Gamma_{23}(1,2,3)=\Gamma_{12}(2,3,1)$ and
$\Gamma_{31}(1,2,3)=\Gamma_{12}(3,1,2)$,
the system is reduced to a single equation for one of the amplitudes, say $\Gamma_{12}$.

In general, $\Gamma_{12}$ depends on
all  variables ($\vec{R}_{i\perp},x_i$), constrained by the relations
$\vec{R}_{1\perp}+\vec{R}_{2\perp}+\vec{R}_{3\perp}=0$, $x_1+x_2+x_3=1$,
but for  a contact kernel it depends only on $(\vec{R}_{3\perp},x_3)$ \cite{tobias}.
Equation (\ref{eq3}) results into:
\begin{equation}\label{eq24}
\Gamma_{12}(\vec{R}_{\perp},x)
=\frac{\lambda}{(2\pi)^3}\int \left[\Gamma_{12}(\vec{R}_{\perp},x)
+2\Gamma_{12}\left(\vec{R'}_{\perp}-x'\vec{R}_{\perp},\;
x'(1-x)\right)\right]
\,
\frac{1}{s'_{12}-M_{12}^2}
\frac{d^2R'_{\perp}dx'}{2x'(1-x')},
\end{equation}
in which
\[s'_{12}=(k'_1+k'_2)^2= \frac{{R'}^2_{\perp}+m^2} {x'(1-x')}  \]
is the effective on shell mass squared of the two-body subsystem,
whereas $M^2_{12}=(k'_1+k'_2-\omega\tau')^2=(p-k_3)^2$
corresponds to its off-shell mass.
It is expressed through $M_3^2,R_{\perp}^2,x$ as
\begin{equation}\label{eq17}
M^2_{12}=(1-x)M_3^2-\frac{R_{\perp}^2+(1-x)m^2}{x}.
\end{equation}
These on- and off-shell masses $s'_{12}$ and $M^2_{12}$ differ from each
other, since $k'_1+k'_2+k_3\neq p$.
On the energy shell, at $\tau'=0$, the value $M^2_{12}$ turns into
$s'_{12}$, what is never reached for a bound state problem.

Since the first term $\Gamma_{12}(\vec{R}_{\perp},x)$ in the integrand does not
depend on the integration variables $\vec{R'}_{\perp},x'$, we can transform (\ref{eq24}) as:
\begin{equation}\label{eq25}
\Gamma_{12}(\vec{R}_{\perp},x)
=\frac{1}{\lambda^{-1}-I(M_{12})}
\frac{2}{(2\pi)^3}\int
\Gamma_{12}\left(\vec{R'}_{\perp}-x'\vec{R}_{\perp},x'(1-x)\right)
\frac{1}{s'_{12}-M_{12}^2}\frac{d^2R'_{\perp}dx'}{2x'(1-x')},
\end{equation}
where
\begin{equation}\label{eq26}
I(M_{12})=\frac{1}{(2\pi)^3}\int \frac{1}{s'_{12}-M_{12}^2}
\frac{d^2R'_{\perp}dx'}{2x'(1-x')}.
\end{equation}
The integral (\ref{eq26}) diverges logarithmically and we
implicitly assume that a cutoff $L$ is introduced.

The value of $\lambda$ is found by solving the two-body problem
with the same zero-range interaction under the
condition that the two-body bound state mass has a fixed value
$M_{2}$. From that we get $\lambda^{-1}= I(M_{2})$ with  $I$ given by
(\ref{eq26}). It also diverges when the
momentum space cutoff $L$ tends to infinity (or, equivalently, the interaction
range tends to zero). However, the difference $\lambda^{-1}-I(M_{12})=
I(M_{2})-I(M_{12})$ which appears in (\ref{eq25}) converges in the limit $L\to\infty$.
The factor $F(M_{12})=
1/[I(M_{2})-I(M_{12})]$ gives the two-body off-shell scattering amplitude,
depending on the off-shell two-body mass $M_{12}$, without any regularization.
For $0\leq M_{12}^2<4m^2$ explicit calculations gives:
$$
F(M_{12})=\frac{8\pi^2}{\frac{\displaystyle{\arctan y_{M_{12}}}}
{\displaystyle{y_{M_{12}}}}
-\frac{\displaystyle{\arctan y_{M_{2}}}}{\displaystyle{y_{M_{2}}}}},
$$
where
$ y_{M_{12}}=\frac{M_{12}}{\sqrt{4m^2-M_{12}^2}}$
and similarly for $y_{M_{2}}$.
If $M_{12}^2<0$, the amplitude obtains the form:
$$
F(M_{12})=\frac{8\pi^2}{\frac{\displaystyle{1}}
{\displaystyle{2y'_{M_{12}}}}\log \frac{\displaystyle{1+y'_{M_{12}}}}
{\displaystyle{1-y'_{M_{12}}}}
-\frac{\displaystyle{\arctan y_{M_{2}}}}
{\displaystyle{y_{M_{2}}}}},
$$
where
$
y'_{M_{12}}=\frac{\sqrt{-M_{12}^2}}{\sqrt{4m^2-M_{12}^2}}.
$

Finally, the equation for the Faddeev amplitude reads:
\begin{equation}\label{eq29a}
\Gamma_{12}(R_{\perp},x)
=F(M_{12})\frac{\displaystyle{1}}{\displaystyle{(2\pi)^3}}
\displaystyle{\int_0^1}
\displaystyle{dx'}
\displaystyle{\int_0^{\infty}}
\frac{\Gamma_{12}\left(R'_{\perp},x'(1-x)\right)\;d^2R'_{\perp}}
{\displaystyle{(\vec{R'}_{\perp}-x'\vec{R}_{\perp})^2+m^2-x'(1-x')M_{12}^2}}.
\end{equation}
The three-body mass $M_3$ enters in this equation through the variable
$M_{12}^2$, defined by (\ref{eq17}).

By replacing  $x'(1-x)\to x'$, equation (\ref{eq29a}) can be transformed into
\begin{equation}\label{eq30}
\Gamma_{12}(R_{\perp},x)
=F(M_{12})\frac{\displaystyle{1}}{\displaystyle{(2\pi)^3}}
\displaystyle{\int_0^{1-x}}
\frac{\displaystyle{dx'}}{\displaystyle{x'(1-x-x')}}\;
\displaystyle{\int_0^{\infty}}
\frac{\displaystyle{d^2R'_{\perp}}}{\displaystyle{{{\cal M}'}^2-M_3^2}}\;
\Gamma_{12}\left(R'_{\perp},x'\right),
\end{equation}
with
\[ {{\cal M}'}^2=\frac{\vec{R'}^2_{\perp}+m^2}{x'}
+\frac{\vec{R}^2_{\perp}+m^2}{x}
+\frac{(\vec{R'}_{\perp}+\vec{R}_{\perp})^2+m^2}{1-x-x'} \]

This equation is the same than the equation (11) from \cite{tobias} except
for the integration limits of ($\vec{R'}_{\perp},x'$) variables.
In \cite{tobias} the integration limits follow from the condition $M_{12}^2>0$.
They read
\begin{equation}\label{FBC}
\int_{m^2\over M_3^2}^{1-x} \left[\ldots\right] dx'
\int_0^{k^{max}_{\perp}} \left[\ldots\right]  d^2R'_{\perp}
\end{equation}
with $k^{max}_{\perp}=\sqrt{(1-x')(M_3^2x'-m^2)}$
and thus implicitly introduce a lower bound on the three-body mass $M_3>\sqrt{2}m$.
The same condition, though in a different relativistic approach, was used in \cite{Noyes}.
The integration limits in (\ref{FBC}) restrict the arguments
of $\Gamma_{12}$ to the domain
\[ {m^2\over M_3^2} \leq x \leq 1 -{m^2\over M_3^2} ,
\quad 0\leq R_{\perp}\leq k_{\perp}^{max}\]
and can be considered as a  method of regularization.
In this case, one no longer deals with the zero-range forces.

Being interested in studying the zero-range interaction,
we do not cut off the variation domain of variables $R_{\perp},x$:
\[0\leq x\leq 1,\quad 0\leq R_{\perp}< \infty \]

The integration limits for these variables
reflect the conservation law of the
four-momenta  in the three-body system and they are automatically fullfilled, as
far as the $\delta^{(4)}$-function in  (\ref{eq3}) is taken into account.
The off-shell variable $M_{12}^2$ may take negative values,
when $R_{\perp}$ and $x$ vary in their proper limits.
Thus, if $M_3^2>m^2$ one has $-\infty \le M_{12}^2 \le (M_3-m)^2$
but if  $M_3^2<m^2$,  $M_{12}^2$ is always negative $-\infty \le M_{12}^2 \le 0$.

We would like to notice that $M_{12}^2$
is not to be confused with the on-shell effective mass squared
$s'_{12}=(k'_1+k'_2)^2$
which is indeed always positive and even $s'_{12}\geq 4m^2$.
As we will see in the next section, this point turns out to be crucial for the appearence of the relativistic collapse.

%%%%%%%%%%%%%%%%%%%%%%%%%%%%%%%%%%%%%%%%%%
\subsection{Results}\label{results}

The results of solving equation (\ref{eq29a}) are
presented in what follows. Calculations were carried out with constituent mass
$m=1$ and correspond to the ground state.
\bigskip
\begin{figure}[hbtp]
\begin{center}
\epsfxsize=8.8cm\subfigure[ ]{\epsffile{M3_M2_F.eps}}\hspace{0.5cm}
\epsfxsize=6.8cm\subfigure[ ]{\epsffile{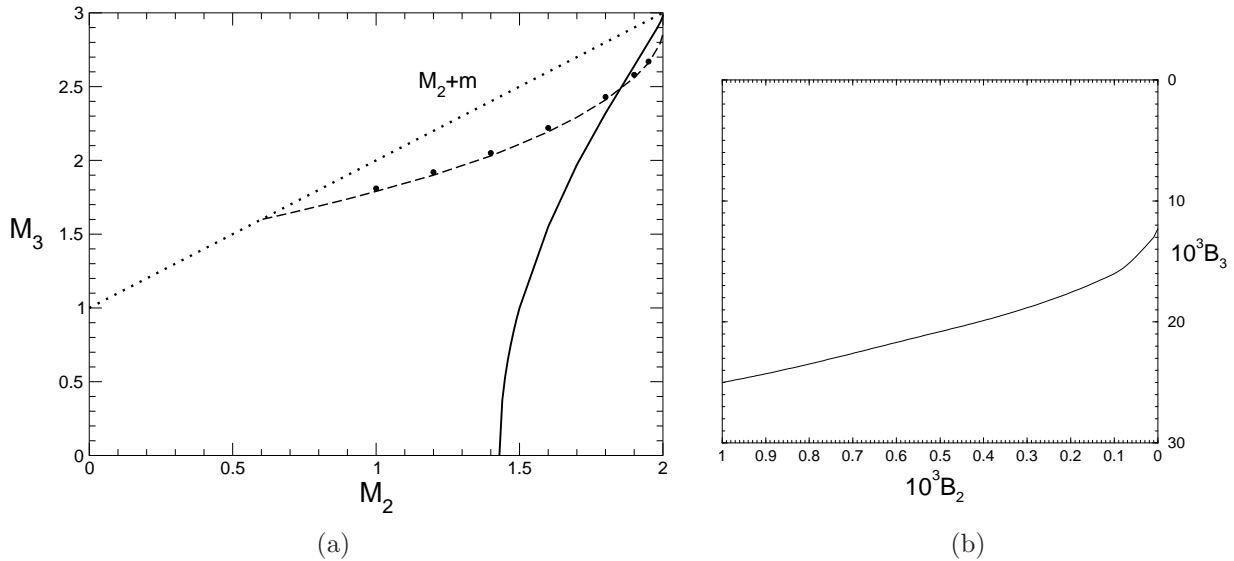}}
\caption{(a) Three-body bound state mass $M_3$ versus the two-body one $M_{2}$ (solid line). Dotted line represents the dissociation limit.
Results  obtained with integration limits (\protect{\ref{FBC}}) are in dash line. Bold dots are taken from \cite{AMF_PRC_95}.
(b) Zoom of the two-body zero binding limit region ($M_2\to2m,B_2=2m-M_2\to0$) corresponding to the solid line only.}\label{Fig_M3_M2}
\end{center}
\end{figure}
We represent in fig.~\ref{Fig_M3_M2}a
the three-body bound state mass $M_3$ as a function of the two-body one $M_2$
(solid line) together with the dissociation limit $M_3=M_2+m$ (dotted line).
The  two-body zero binding limit $B_2=2m-M_2\to0$ is magnified in fig.~\ref{Fig_M3_M2}b. In this limit
the three-boson system has a binding energy  $B_3^{(c)}\approx0.012$.

When $M_2$ decreases, the three-body mass $M_3$  decreases very
quickly and vanishes at the two-body mass value $M_{2}= M_2^{(c)} \approx 1.43$.
Whereas the meaning of collapse as used in the Thomas paper \cite{thomas} implies unbounded
nonrelativistic binding energies and cannot be used here,
the zero bound state mass $M_3=0$ constitutes its relativistic counterpart.
Indeed, for two-body masses below the critical value $M_2^{(c)}$, the three-body system  no longer exists.
\begin{figure}[!ht]
\begin{center}
\mbox{\epsfxsize=7.cm\epsffile{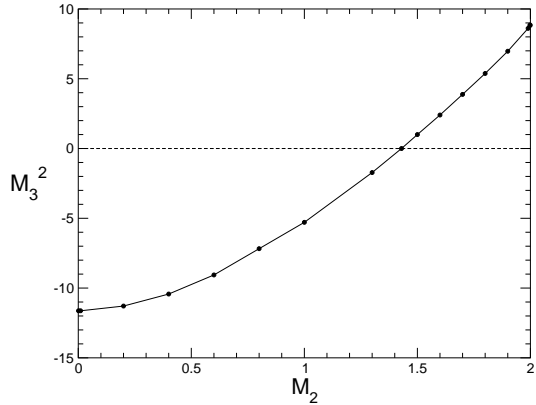}}
\caption{Three-body bound state mass squared $M_3^2$ versus $M_{2}$.}
\label{Fig_M32_M2}
\end{center}
\end{figure}

The results corresponding to integration limits (\ref{FBC})
are included in fig.~\ref{Fig_M3_M2}a (dash line) for comparison.
Values given in \cite{tobias} were not fully converged. 
They have been corrected in \cite{AMF_PRC_95} and are indicated by dots.
In both cases the repulsive relativistic effects produce a natural cutoff in
equation (\ref{eq29a}), leading to a finite spectrum and -- in the Thomas sense
-- an absence of collapse, like it was already found in \cite{Noyes}.
However,  solid and dash curves strongly differ from each other, even in the zero binding limit.

We would like to remark that for $M_2\leq M_2^{(c)}$, equation (\ref{eq29a}) posses
square integrable solutions with negative values of $M_3^2$. They have no
physical meaning but $M_3^2$ remains finite in all the two-body mass range
$M_2\in[0,2]$. The results of $M_3^2$ are given in figure \ref{Fig_M32_M2}.
When $M_{2}\to 0$,  $M_3^2$ tends to $\approx -11.6$.

It is also worth noticing that the critical value of the two-body bound state mass $M_2^{(c)}$ 
as well as the three-body binding energy $B_3^{(c)}$ are universal quantities for bosonic systems.
$M_2^{(c)}=1.43 \,m$ represents the maximal two-body binding energy $B_2=2m-M_2^{(c)}=0.57\,m$ compatible with the existence of 3-boson bound states with mass $M_3= 0$ ($B_3=3\,m$). $B_3^{(c)}=0.012\,m$ represents the minimal binding energy that a three-boson system can have when two-body binding energy $B_2=0$ ($M_2=2\,m$).

%%%%%%%%%%%%%%%%%%%%%%%%%%%
\section{Two-fermion system with Yukawa interaction}
\subsection{States with $J=0$}

So far we have considered the behavior of  the  three-boson relativistic bound system and its critical stability
depending on the two-body binding energy. The conclusion are valid for the zero-range interactions, considered as input for the two-body sector, 
and we have supposed that the particles were spinles.

Now we will study a system of two fermions -- spin 1/2 particles -- with more sophisticated
interaction, resulting from spinless mesons exchange with mass $\mu$. This model traces back to the very origin of the nuclear forces theory proposed by Yukawa. The interaction Lagrangian reads:
$$
{\cal L}^{int}=g\;\bar{\psi}\psi\phi
$$
Let us consider first the case of zero total angular momentum $J=0$.
We denote the fermion momenta as $\vec{k}_1$, $\vec{k}_2$. It is convenient to analyze the wave function in the reference frame where $\vec{k}_1=\vec{k}_2=0$. Then the two-fermion wave function depends on the relative momentum 
$\vec{k}=\vec{k}_1=-\vec{k}_2$ and on the spin projections of each fermion $\sigma_1,\sigma_2=\pm 1/2$. 
Relative to the spin projections, it is a $2\times 2$ matrix which has the following general form \cite{2ferm}:
\begin{equation}\label{eq0} 
\psi(\vec{k},\vec{n})=\frac{1}{\sqrt{2}}\left(f_1+ \frac{i\vec{\sigma}\cd [\vec{k}\times \vec{n}]}
{\sin\theta}f_2\right),
\end{equation}
where $\vec{\sigma}$ are the Pauli matrices, $\vec{n}=\vec{\omega}/ \omega_0$ and $\theta$ is the angle between
$\vec{k}$ and $\vec{n}$. One cannot construct any other independent structures in addition to those  appearing in (\ref{eq0}). Therefore the $2\times 2$ matrix $\psi(\vec{k},\vec{n})$ contains only two independent matrix elements 
or, correspondingly, it is determined by the two coefficients $f_1,f_2$ of the independent structures. 
The normalization condition has the form:
$$
\frac{m}{(2\pi)^3}\int (f_1^2+ f_2^2) {d^3k\over\varepsilon_k}=1,
\quad \varepsilon_k=\sqrt{m^2+k^2}.
$$
The equation for the wave function is reduced to a system of two coupled equations for $f_{1,2}$:
\begin{eqnarray}\label{eq10a}  
&&\left[4(k^2 +m^2)-M^2\right] f_1(k,\theta) 
\nonumber\\
&&=-\frac{m^2}{2\pi^3} \int
\left[K_{11}(k,\theta;k',\theta')
f_1(k',\theta')+K_{12}(k,\theta;k',\theta')
f_2(k',\theta')\right]\frac{d^3k'}{\varepsilon_{k'}},
\nonumber\\
&&\left[4(k^2 + m^2)-M^2\right] f_2(k,\theta) 
\nonumber\\
&&=-\frac{m^2}{2\pi^3} \int \left[K_{21}(k,\theta;k',\theta')
f_1(k',\theta')+K_{22}(k,\theta;k',\theta')
f_2(k',\theta')\right]\frac{d^3k'}{\varepsilon_{k'}}
\end{eqnarray}
with the kernels:
\begin{equation}\label{nz1} 
K_{ij}=\int_0^{2\pi}{\kappa_{ij} \over
(K^2+\mu^2)m^2\varepsilon_k \varepsilon_{k'}}{d\phi'\over 2\pi}, 
\label{nz1_k} 
\end{equation}
where
\begin{eqnarray}\label{k4} 
K^2 &=& k^2+k'^2
-2kk'\left(1+\frac{(\varepsilon_k -\varepsilon_{k'})^2}
{2\varepsilon_k\varepsilon_{k'}}\right)\cos\theta\cos\theta'
-2k k'\sin\theta \sin\theta'\cos\phi'
\nonumber\\
&&+\left(\varepsilon_{k}^2
+\varepsilon_{k'}^2-\frac{1}{2}M^2\right)                           
\left|\frac{k\cos\theta}{\varepsilon_{k}}                                
-\frac{k'\cos\theta'}{\varepsilon_{k'}}\right|                                 
\end{eqnarray}
Here $\phi'$ is the
azimuthal angle between $\vec{k}$ and $\vec{k}\,'$ in the plane orthogonal
to $\vec{n}$
and 
$$\cos\theta=\cos\vec{n}\cd\vec{k}/k, \quad
\cos\theta'=\cos\vec{n}\cd\vec{k}\,'/k'.$$
The explicit expressions for $\kappa_{ij}$ is given by \cite{2ferm,mck4}:
\begin{eqnarray}\label{eqap1}
\kappa_{11}&=&-\alpha\pi
\left[2 k^2 k'^2+3k^2 m^2+3k'^2 m^2+4 m^4
-2 k k'\varepsilon_k \varepsilon_{k'} \cos\theta \cos\theta' \right.
\nonumber\\
&&\left.- k k' (k^2 + k'^2 + 2 m^2)
\sin\theta \sin\theta' \cos\phi'\right],\nonumber\\
\kappa_{12}&=&-\alpha\pi m
(k^2 - k'^2) \left(k'\sin\theta' + k\sin\theta\cos\phi' \right),
\nonumber\\
\kappa_{21}&=&-\alpha\pi m
(k'^2 - k^2) \left(k\sin\theta + k'\sin\theta'\cos\phi' \right),
\nonumber\\
\kappa_{22}&=&-\alpha\pi
\left[\left(2 k^2 k'^2+3k^2 m^2+3k'^2 m^2+4 m^4
- 2 k k' \varepsilon_k\varepsilon_{k'}
\cos\theta \cos\theta'\right)\cos\phi'\right.
\nonumber\\
&&\left.-k k'(k^2 + k'^2 + 2 m^2) \sin\theta\sin\theta'\right],	
\end{eqnarray}
 where we denote $\alpha=g^2/(4\pi)$.

%%%%%%%%%%%%%%%%%%%%%%%%%%%%%%%%%%%%%%%%%%%%%%%%%%%
\subsection{Asymptotical behavior of the kernels}\label{asympt}

The r.h.-sides of equations (\ref{eq10a}) contain the integrals over $k'$ in infinite limits. The existence of a finite
solution  depends critically on the behavior of the kernels $K_{ij}$ at large momenta.
For the kernels (\ref{eqap1})  determining the $J=0$ state, we get the following  leading terms:
\begin{eqnarray}\label{as1}
&&K_{11}\propto
\left\{
\begin{array}{ll}
\frac{\displaystyle{1}}{\displaystyle{k}},
&\mbox{if $k\to\infty$, $k'$ fixed}
\\
&
\\
\frac{\displaystyle{1}}{\displaystyle{k'}},
&\mbox{if $k'\to\infty$, $k$ fixed}
\end{array}
\right.
\nonumber\\
&&K_{12}\propto
\left\{
\begin{array}{ll}
\frac{\displaystyle{1}}{\displaystyle{k}},
&\mbox{if $k\to\infty$, $k'$ fixed}
\\
c_{12},
&\mbox{if $k'\to\infty$, $k$ fixed}
\end{array}
\right.
\nonumber\\
&&K_{21}\propto
\left\{
\begin{array}{ll}
c_{21}=c_{12},
&\mbox{if $k\to\infty$, $k'$ fixed}   \\
\frac{\displaystyle{1}}{\displaystyle{k'}},
&\mbox{if $k'\to\infty$, $k$ fixed}
\end{array}
\right.
\nonumber\\
&&K_{22}=
\left\{
\begin{array}{ll}
c_{22},
&\mbox{if $k\to\infty$, $k'$ fixed},   \\
c'_{22}, &\mbox{if $k'\to\infty$, $k$ fixed}
\end{array}
\right.
\end{eqnarray}
In the above equations the coefficients $c_{12}=c_{21}$, $c_{22}$ and $c'_{22}$ depend  on $\theta,\theta'$.
The coefficients $c_{22},c'_{22}$ are positive:
\begin{equation}\label{as7}
c_{22}=\frac{\alpha\pi\sin\theta\sin\theta'}
{m(1+\cos\theta)(\varepsilon_{k'}-k'\cos\theta')}>0,
\end{equation}
and $c'_{22}$ is obtained form $c_{22}$ by the replacement
$k'\rightarrow k,\theta\leftrightarrow \theta'$. 

Note that the second iteration of the kernel $K_{11}$ 
converges at $k'\to\infty$:
$$
\int^L K_{11}G_0K_{11}\frac{d^3k'}{\varepsilon_{k'}}\propto
\int^L \frac{1}{k'}\frac{1}{k'^2}\frac{1}{k'}\frac{k'^2dk'}{k'}=
\int^L \frac{dk'}{k'^3}\propto const.
$$
Here $G_0\propto 1/k'^2$ is the intermediate propagator.
The integrals 
\[ \int K_{21}G_0K_{11}d^3k'/\varepsilon_{k'}  \qquad,\qquad  \int K_{11}G_0K_{12}d^3k'/\varepsilon_{k'}\]
are also convergent,  whereas the the second iteration of the kernel $K_{22}$  diverges logarithmically:
$$
\int^L K_{22}G_0K_{22}\frac{d^3k'}{\varepsilon_{k'}}\propto
\int^L const \frac{1}{k'^2}const \frac{k'^2dk'}{k'}=
\int^L\frac{dk'}{k'}\propto \log(L).
$$
The integrals 
\[ \int K_{12}G_0K_{22}d^3k'/\varepsilon_{k'} \qquad,\qquad  \int K_{22}G_0K_{21}d^3k'/\varepsilon_{k'}\]
also diverge logarithmically. This is a manifestation of the logarithmical divergence of  the box fermion diagram in LFD.

In the domain where both $k,k'$ tend to infinity, but the ratio $k'/k=\gamma$ is fixed, we find for $K_{11}$:
\begin{equation}\label{eqn16}
K_{11}=-\frac{2\pi^2\alpha'}{m}\left\{
\begin{array}{ll}
\sqrt{\gamma}A_{11}(\theta,\theta',\gamma),
& \mbox{if $\gamma\leq 1$}
\\
\frac{\displaystyle{A_{11}(\theta,\theta',1/\gamma)}}
{\displaystyle{\sqrt{\gamma}}},
& \mbox{if $\gamma\geq 1$}
\end{array}\right.
\end{equation}
with the function $A_{11}(\theta,\theta',\gamma)$:
\begin{equation}\label{eq14a}
A_{11}(\theta,\theta',\gamma)=
\frac{1}{\sqrt{\gamma}}\int_0^{2\pi}\frac{d\phi}{2\pi}
\frac{2\gamma(1-\cos\theta\cos\theta')-
(1+\gamma^2)\sin\theta\sin\theta'\cos\phi}
{(1+\gamma^2)(1+|\cos\theta-cos\theta'|-\cos\theta\cos\theta')
-2\gamma\sin\theta\sin\theta'\cos\phi},
\end{equation}
where we set $\alpha'=\alpha/(2m\pi)$. In eq. (\ref{eqn16}) we extracted for convenience the factor $\sqrt{\gamma}$. In the limit $\gamma\to 0$ $A_{11}$ has the behavior  $A_{11}(\theta,\theta',\gamma)\propto \sqrt{\gamma}$.

In the same domain, the kernel $K_{22}$ also has asymptotic (\ref{eqn16}) with the corresponding function $A_{22}$ given by:
\begin{equation}\label{eq14a_1}
A_{22}(\theta,\theta',\gamma)=
-\frac{1}{\sqrt{\gamma}}\int_0^{2\pi}\frac{d\phi}{2\pi}
\frac{(1+\gamma^2)\sin\theta\sin\theta'-
2\gamma(1-\cos\theta\cos\theta')\cos\phi}
{(1+\gamma^2)(1+|\cos\theta-cos\theta'|-\cos\theta\cos\theta')
-2\gamma\sin\theta\sin\theta'\cos\phi}.
\end{equation}
In the limit $\gamma\to 0$ this function has the behavior  $A_{22}(\theta,\theta',\gamma)\propto -1/\sqrt{\gamma}$.

Comparing the above formulas, we see that the dominating kernel is 
$K_{22}$. It does not decreases in any direction of the $(k,k')$ plane,
whereas in the domain $k\to\infty$, $k'$ fixed, and vice versa, 
the kernels $K_{11}$ decrease. 
In the domain $k'/k=\gamma$ fixed, $k\to\infty$, both kernels do not decrease, but $K_{22}$ is proportional to the unbounded function $A_{22}$.

%%%%%%%%%%%%%%%%%%%%%%%%%%%%%%%%%%%%%%%%%%%%%%%%%%%%%
\subsection{The cutoff dependence of the binding energy}\label{depend}

We are now  in position to investigate the stability of the bound states.
To disentangle the two different sources of collapse, we will first consider the one channel problem  for the component $f_1$ with the kernel $K_{11}$. 
We remove the second equation from (\ref{eq10a}) and deal with the single equation:
\begin{equation}\label{as2_1}  
\left[4(\vec{k}\,^2 +
m^2)-M^2\right] f_1(k,z) =-\frac{m^2}{2\pi^3} \int
K_{11}(k,z;k',z')f_1(k')\frac{d^3k'}{\varepsilon_{k'}}.
\end{equation}

Our further analysis is based on the collpas condition found by Smirnov
\cite{smirnov}. It is obtained by  analyzing the asymptotic of eq. (\ref{as2_1}) with the kernel represented by eq. (\ref{eqn16}). 
The solution is searched in the form 
\begin{equation}\label{as2_3}
f_1(k,z)\propto \frac{f(z)}{k^{2+\beta}}.
\end{equation}
In eq. (\ref{as2_1}) one should make the replacement of variables 
$k'=\gamma k$ and take the limit $k\to \infty$. Provided the kernel 
$K_{11}(k,k'=const)$ decreases like $1/\sqrt{k}$ or faster, one gets:
$$
4k^2 \frac{f(z)}{k^{2+\beta}}=-\frac{m^2}{2\pi^3} \int 
K_{11}(k,z;k\gamma,z')\frac{f(z')}{(k\gamma)^{2+\beta}}
\frac{ k^3 \gamma^2 d\gamma 2\pi dz'}{k\gamma}.
$$
Splitting the  integral in two terms:
\[ \int_0^{\infty}\ldots d\gamma =\int_0^{1}\ldots d\gamma+
\int_1^{\infty}\ldots d\gamma  , \]
making in the second term the substitution $\gamma=1/\gamma'$ and substituting here the kernel (\ref{eqn16}), we obtain the equation:
\begin{equation}\label{cut1}
f(z)=2 m\alpha'\int_0^1  dz' f(z') 
\int_0^1 d\gamma \frac{A_{11}(\gamma,z,z')}{\sqrt{\gamma}}
\cosh(\beta\log(\gamma))
\end{equation}
Using the symmetry relative to $z\to 1-z$, we replaced the integral 
$\int_{-1}^1\ldots dz'$ by $2\int_0^1\ldots dz'$.

In the above equations we neglected the binding energy, supposing that it
is finite. For given $\alpha'$ the equation (\ref{cut1}) gives the value
of $\beta$, determining the wave function asymptotic (\ref{as2_3})
for the solution with finite energy.
The function $\cosh(\beta\log(\gamma))$ in (\ref{cut1})
has minimum at $\beta=0$.
When the factor $\alpha'$ in (\ref{cut1}) increases, this is 
compensated by decrease of $\cosh(\beta\log(\gamma))$, so
the value of $\beta$ is approaching to 0. The 
maximal, critical value of $\alpha'$ is achieved when $\beta=0$.
So, if we solve the eigenvalue equation \cite{smirnov}:
\begin{equation}\label{cut2}
\int_0^1 H(z,z')f(z')d z'=\lambda f(z), 
\quad \mbox{with}\quad 
H(z,z')=2\int_0^1\frac{A_{11}(\gamma,z,z')}{\sqrt{\gamma}}d\gamma 
\end{equation}
then the critical value of $\alpha'$ is related to $\lambda$ as
$\alpha'_c=\frac{1}{m\lambda}$, that gives for the coupling constant 
in the Yukawa model $\alpha=g^2/(4\pi)=2\pi m\alpha'$ the following 
critical value:
\begin{equation}\label{cut3} 
\alpha_c=\frac{2\pi}{\lambda}.
\end{equation}
Note that if $A_{11}(\gamma,z,z')=A(\gamma)$ does not depend on $z,z'$, 
one gets \cite{smirnov}:
 \begin{equation}\label{cut4}
\alpha'_c=\frac{1}{2m\int_0^1
\frac{\displaystyle{A(\gamma)}}{\displaystyle{\sqrt{\gamma}}}d\gamma}.
\end{equation}

For the potential $V(r)=-\alpha'/r^2$ one can find  $A(\gamma)=1$ and one gets the well known value $\alpha'_c=1/(4m)$  \cite{ll}. 
In \cite{mck3}
we have estimated $\alpha_c=\pi$ by majorating the kernel $A_{11}$ by
$A_{11}=\sqrt{\gamma}$. Substitution of this function $A_{11}$ into 
eq. (\ref{cut4}) reproduces this result.

Solving eq. (\ref{cut2}) numerically with the function $A_{11}(\gamma,z,z')$
given by eq. (\ref{eqn16}), we found the only eigenvalue:
$$\lambda=1.748$$
that gives by eq. (\ref{cut3}):
\[ \alpha_c=3.594  \quad \Longleftrightarrow \quad g_c= \sqrt{4\pi \alpha_c}=  6.720 \] 
in agreement with our numerical estimations \cite{mck3}.

In the two-channel problem, the kernel  dominating in  asymptotic is $K_{22}$.
In the case $J=0$ it is positive and corresponds to repulsion. Because of that,
this channel does not lead to any collapse. This repulsion cannot prevent from
the collapse in the first channel (for enough large $\alpha$), since due to
coupling between two  channels the singular potential in the channel 1 "pumps
out" the wave function from the channel 2 into the channel 1. So, in the
coupled equations system  (\ref{eq10a}) the situation with the cutoff
dependence is the same as for one channel.
A similar analysis of what we detailed in the one channel case, provided us  the critical value of the coupling constant \cite{mck3,mck4}
\begin{equation}\label{CCC_LFD}
 \alpha_c=3.723  \quad \Longleftrightarrow \quad g_c= \sqrt{4\pi \alpha_c}=  6.840 
\end{equation} 
 
The critical stability of the Yukawa model has been also considered in the framework of the Bethe-Salpeter equation \cite{Salpeter:1951sz}.
By using the methods developed in the previous section we have found \cite{Carbonell:2010zw,Carbonell:2010tz,CKdS_2013} similar results of 
what we have obtained in the Light-Front dynamics.
There very existence of  a critical coupling constant for the J=0 state was confirmed, although with slightly different numerical value:
\begin{equation}\label{CCC_BS}
 \alpha_c=\pi  \quad \Longleftrightarrow \quad g_c= 2\pi 
 \end{equation}
 to be compared with (\ref{CCC_LFD}).

%%%%%%%%%%%%%%%%%%%
\subsection{States with $J=1$}

In general, the wave function of the $J=1$ state is determined by six independent structures  \cite{ckj1}.
It turns out that   the following operator commutes with the kernel:
\begin{equation}\label{ac6}                                                     
A^2=(\vec{n}\cd\vec{J})^2.
\end{equation}
Since $A^2$ is a scalar, it commutes also with $\vec{J}$.
Therefore, in addition to $J,J_z$, the solutions are labeled by $a$:
\begin{equation}\label{eqf1}
A^2\vec{\psi}^{a}(\vec{k},\vec{n})=
a^2\vec{\psi}^{a}(\vec{k},\vec{n}).
\end{equation}
Though the wave function for $J=1$ is determined by
six components,  the equation system is split in
two subsystems with $a=0$ and $a=1$, containing 2 and 4 equations respectively \cite{2ferm}.

The function $\vec{\psi}^0$ corresponding to $J=1$, $a=0$
has the following general decomposition:
\begin{equation}\label{eq4a} 
\vec{\psi}^0(\vec{k},\vec{n})=\sqrt{\frac{3}{2}}\left\{g^{(0)}_1
\vec{\sigma}\cd\hat{\vec{k}}
+
g^{(0)}_2 \frac{\vec{\sigma}\cd(\hat{\vec{k}}\cos\theta-\vec{n})}
{\sin\theta}\right\}\vec{n},
\end{equation} 
where $\hat{\vec{k}}$ denotes the unit vector $\hat{\vec{k}}=\vec{k}/k$.
Since it corresponds to $J^{\pi}=1^+$, it is a pseudovector. 
Since $\vec{n}$ is a true vector ($J^{\pi}=1^-$), it should be multiplied by
a pseudoscalar. We can construct two pseudoscalars only: 
$\vec{\sigma}\cd \hat{\vec{k}}$ and $\vec{\sigma}\cd \hat{\vec{n}}$, what 
gives two terms. The particular structures in (\ref{eq4a}) are constructed in such a way to be orthogonal and normalized to 1.

The function $\vec{\psi}^1$ satisfies the orthogonality 
condition $\vec{\psi}^1\cd \vec{n}=0$. To satisfy this condition, 
it is convenient to introduce the vectors orthogonal to $\vec{n}$:
$$
\hat{\vec{k}}_\perp= \frac{\hat{\vec{k}}-\cos\theta\vec{n}}{\sin\theta},
\quad
\vec{\sigma}_\perp= \vec{\sigma}-(\vec{n}\cd \vec{\sigma})\vec{n}.
$$
Then the function $\vec{\psi}^1$ obtains
the following general form:
\begin{equation}\label{eq4b} 
\vec{\psi}^1(\vec{k},\vec{n})=
g^{(1)}_1\frac{\sqrt{3}}{2}\vec{\sigma}_\perp
+g^{(1)}_2\frac{\sqrt{3}}{2}\left(2\hat{\vec{k}}_\perp 
(\hat{\vec{k}}_\perp \cd \vec{\sigma}_\perp)-\vec{\sigma}_\perp\right)
+g^{(1)}_3\sqrt{\frac{3}{2}}\hat{\vec{k}}_\perp (\vec{\sigma}\cd \vec{n})
+g^{(1)}_4\sqrt{\frac{3}{2}}i[\hat{\vec{k}}\times \vec{n}]
\end{equation}

In summary, the system of six equations for the J=1 state is split in two subsystems: two equations for $a=0$  and four for $a=1$.
The subsystem for $a=0$ has the same structure than (\ref{eq10a} ) with different kernels $K^{(J=1)}_{ij}$. 
The asymptotic of  the kernel
$K^{(J=1)}_{22}$  is the same than $-K^{(J=0)}_{22}$: it is negative and
corresponds to attraction. The integral (\ref{cut2}) for the kernel $H(z,z')$
with the function $A_{22}$ given by (\ref{eq14a_1}) diverges logarithmically.
Therefore it results in a collapse for any value of the
coupling constant. This result coincides with conclusion of the paper \cite{glazek1}.

%%%%%%%%%%%%%%%%%%%%%%%%%%%%%%%%%%%%%%%%%%%%%%
\subsection{Numerical results}\label{num}

\begin{figure}[hbtp]
\begin{minipage}[t]{77mm}
\begin{center}
\mbox{\epsfxsize=7.5cm\epsfysize=6.0cm\epsffile{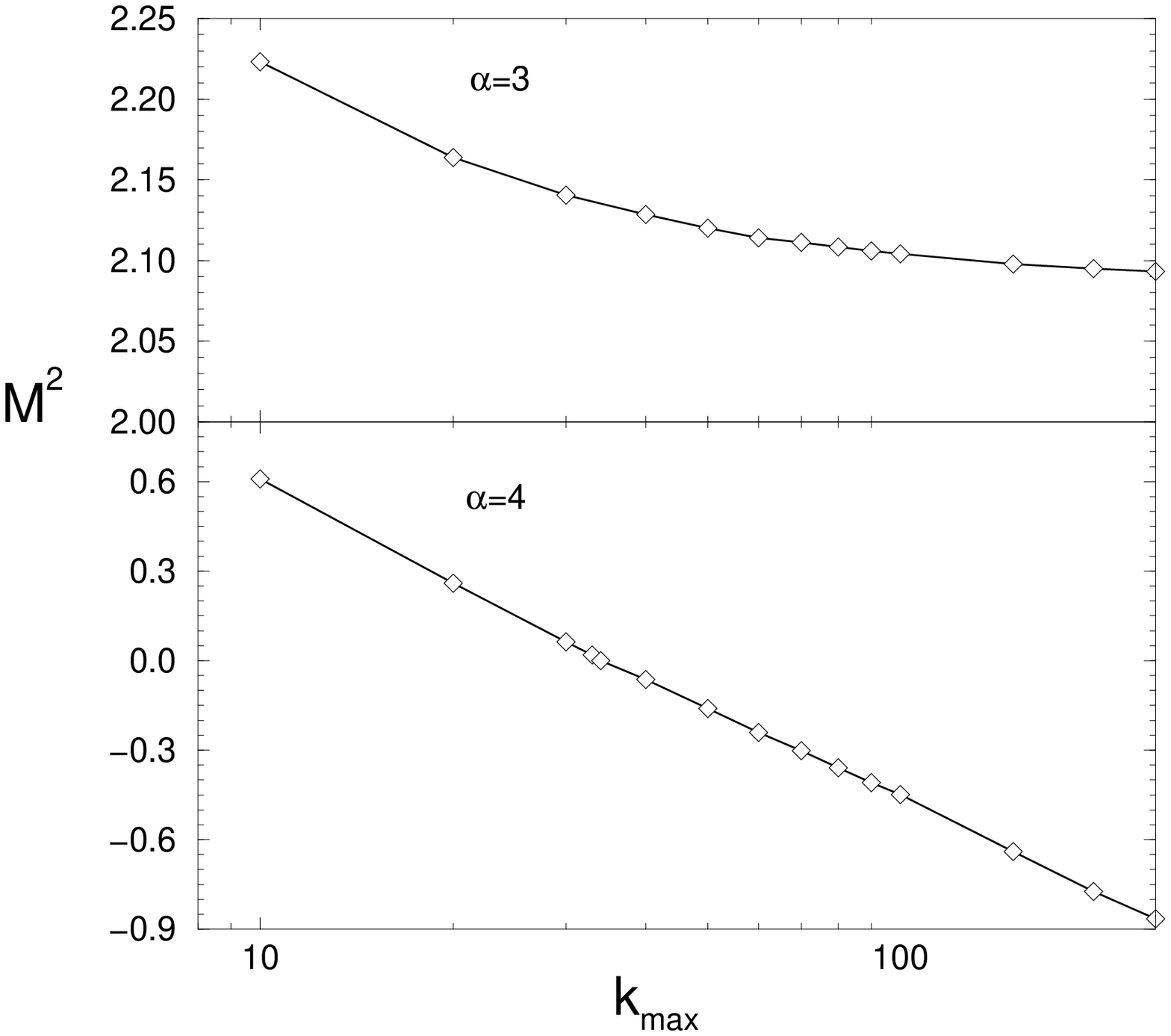}}
\end{center}
\caption{Cutoff dependence of the binding energy in the $J=0$ % or $(1+,2-)$
state,
in the one-channel problem ($f_1$), for two fixed values of the coupling 
constant below and above the critical value.}\label{B_kmax}
\end{minipage}
\hspace{0.5cm}
\begin{minipage}[t]{77mm}
\begin{center}
\mbox{\epsfxsize=7.5cm\epsfysize=7.5cm\epsffile{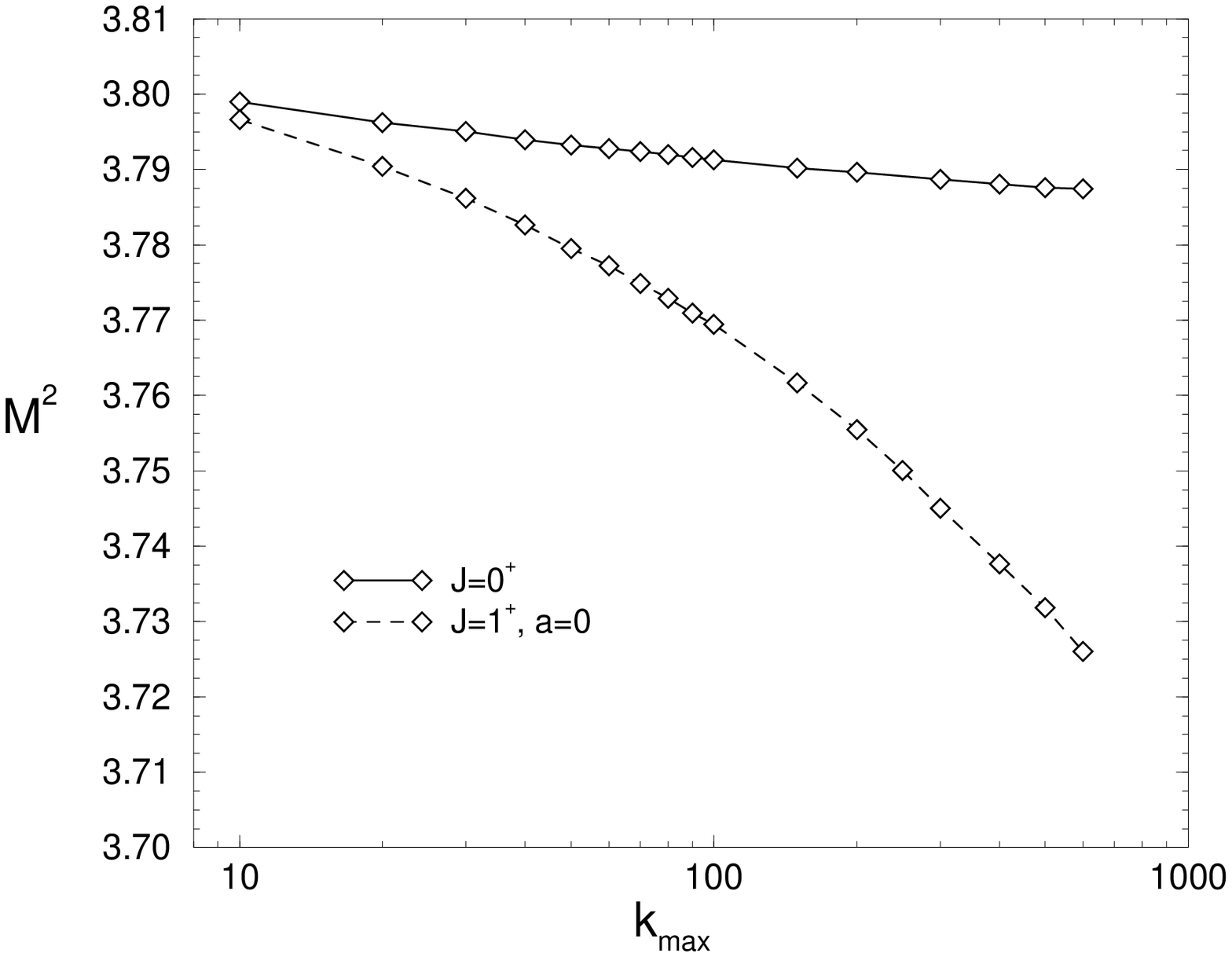}}
\end{center}
\caption{Cutoff dependence of the binding energy, for $J=0$  % $(1+,2-)$ 
and 
$J=1,J_z=0$ %$(1-,2+)$ 
states, in full (two-channel) problem, 
for $\alpha=1.184$.}
\label{alpha_kmax_500}
\end{minipage}
\end{figure}

The preceding analysis are confirmed by several numerical calculations.
In all what follows,  the constituent masses were taken equal to $m$=1 and the mass of the exchanged scalar $\mu$=0.25.

Let us first present the results given by the one channel problem: a single equation for $f_1$ with 
kernel $K_{11}$ in the $J=0$ case. We have plotted in figure \ref{B_kmax} the
mass square $M^2$ of the two fermion system  as a function of the cutoff
$k_{max}$ for two fixed values of the coupling constant below and above  the
critical value. In our calculations  the cutoff appears directly as the maximum
value $k_{max}$  up to which the integrals in (\ref{eq10a}) are performed. One
can see two dramatically different behaviors depending on the value of the
coupling constant $\alpha$. For $\alpha=3$, i.e.  $\alpha<\alpha_{c}=3.594$,
the
result is convergent. For $\alpha=4$, i.e. $\alpha>\alpha_{c}$, the result is
clearly divergent. $M^2$ decreases logarithmically as a function of $k_{max}$
and becomes even negative.  This property is due only to the large $k$ behavior of
$K_{11}$.  Though the negative values of $M^2$ which appear in fig.~\ref{B_kmax}
 are physically meaningless, they are formally  allowed by the equations (\ref{eq10a}). 
The first
degree of $M$ does not enter neither in the equation nor in the kernel, and
$M^2$  crosses zero without any singularity.  The value of the critical
$\alpha$ does not depend on the exchange mass $\mu$. For $\mu\ll m$, e.g.
$\mu\approx0.25$, its existence is not relevant in describing physical states  
since any solution with positive $M^2$, stable relative to cutoff, corresponds
to $\alpha<\alpha_c$. For $\mu\sim m$ one can reach the critical $\alpha$  for
positive, though small values of $M^2$.

We consider now the full Yukawa problem as given by  the two coupled equations
(\ref{eq10a}).  In figure \ref{alpha_kmax_500} are displayed the variations of
$M^2$ for $J=0$ 
and $J=1,J_z=0$ 
states as a function
of the cutoff $k_{max}$. The value of the coupling constant for both $J$  is
$\alpha_c=1.184$, the same that in fig. 2 of \cite{glazek1}, below the critical
value. Our numerical values are in agreement with the results  for the cutoff
$\Lambda \leq 100$  presented in this figure \cite{glazek1}, but our
calculation at larger $k_{max}$ leads to different conclusion for the $J=0$
state. We first notice a qualitatively different behavior of the two states. In
what concerns  $J=0$, the numerical results become more flat when $k_{max}$
increases,  -- with less than a 0.5\% variation in $M^2$ when changing
$k_{max}$ between $k_{max}$=10 and 300. This strongly suggests a convergence.
We thus conclude to the stability of the state with $J=0$, as expected from our
analysis in sect. \ref{depend}.

On the contrary, for $J=1,J_z=0$ the value of  $M^2(k_{max})$ continues to
decrease  faster than logarithmically and indicates, -- as found in
\cite{glazek1}, -- a collapse. As mentioned above, the asymptotic of the 
$K^{(J=1)}_{22}$ kernel is the same as the $K^{(J=0)}_{22}$ one but with an
opposite sign, i.e. it is attractive, what leads to instability for any value
of $\alpha$. The same result was found when solving the $J=0$ equations with
the opposite sign of $K^{(J=0)}_{22}$.

%%%%%%%%%%%%%%%%%%%%
\subsection{Positronium}\label{positronium}

We applied our method to the positronium system  in the $J=0^-$, a bound state of electron and positron which exists in nature.
We consider this important  application in more detail.

The wave  function is again determined by two components and has the form (\ref{eq0}). 
The negative parity of
the state comes from the intrinsic positron parity so that the
corresponding kernels are those of the $J^{\pi}=0^+$ two-fermion
system. They were derived, for the  Feynman gauge  in \cite{2ferm} (eqs. (A8) in appendix A). They have the 
form (\ref{nz1}) with the following values $\kappa_{ij}$ instead of eqs. (\ref{eqap1}) for the scalar case:

\begin{eqnarray}
\kappa_{11}&=&
-2\pi\alpha (4k^2{k'}^2+3m^2(k^2+{k'}^2)+2m^4)\\
\kappa_{12}&=&\phantom{-}2\pi\alpha m k' (k^2-{k'}^2)\sin\theta'\nonumber\\
\kappa_{21}&=&-2\pi\alpha m k(k^2-{k'}^2)\sin\theta\label{vkern}\nonumber\\
\kappa_{22}&=&-2\pi\alpha[kk'(k^2+{k'}^2+2m^2)\sin\theta\sin\theta'+2\epsilon_{k}\epsilon_{k'} (\epsilon_{k}\epsilon_{k'}+
kk'\cos\theta\cos\theta')\cos\phi']
\nonumber
\end{eqnarray}
Following sect. \ref{asympt}, we substitute $k'=\gamma k$ and take the limit $k\to\infty$. The non-diagonal kernels tend to zero, whereas for $K_{11}$ and $K_{22}$ we reproduce (\ref{eqn16})  with the following kernels 
$A(\theta,\theta',\gamma)$:

\begin{equation}\label{A11pos}
A_{11}(\theta,\theta',\gamma)=
8\sqrt{\gamma}\int_0^{2\pi}\frac{d\phi}{2\pi}
\frac{1}
{(1+\gamma^2)(1+|\cos\theta-cos\theta'|-\cos\theta\cos\theta')
-2\gamma\sin\theta\sin\theta'\cos\phi},
\end{equation}
\begin{equation}\label{A22pos}
A_{22}(\theta,\theta',\gamma)=
\frac{2}{\sqrt{\gamma}}\int_0^{2\pi}\frac{d\phi}{2\pi}
\frac{(1+\gamma^2)\sin\theta\sin\theta' +2\gamma(1+\cos\theta\cos\theta')\cos\phi}
{(1+\gamma^2)(1+|\cos\theta-cos\theta'|-\cos\theta\cos\theta')
-2\gamma\sin\theta\sin\theta'\cos\phi},
\end{equation}
where we denote $\alpha'=\alpha/(2m\pi)$. 

At $\gamma\to 0$ $A_{22}$ has the behavior: 
$A_{22}(\theta,\theta',\gamma)\propto +1/\sqrt{\gamma}$ (compare with $A_{22}(\theta,\theta',\gamma)\propto -1/\sqrt{\gamma}$ in eq. (\ref{eq14a_1}) for Yukawa model).

As discussed at the end of sect. \ref{depend}, the behavior $A_{22}(\theta,\theta',\gamma)\propto -1/\sqrt{\gamma}$
corresponds to repulsion, hence for positronium with $A_{22}(\theta,\theta',\gamma)\propto +1/\sqrt{\gamma}$ we have attraction.  The integral  (\ref{cut2}) diverges and the spectrum is unbounded from below.

\begin{figure}[htbp]
\begin{center}
\mbox{\epsfxsize=8cm\epsffile{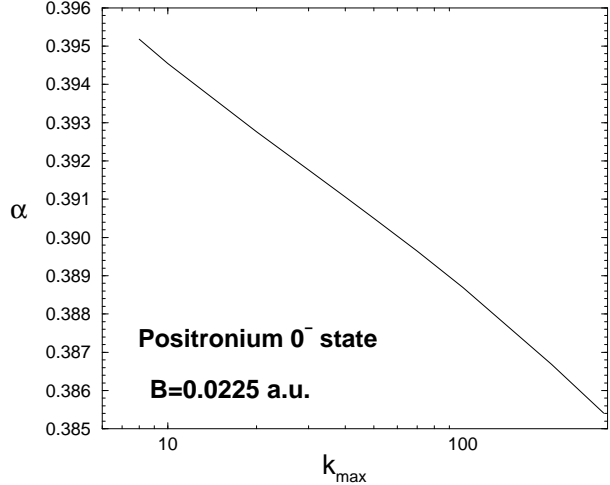}}
\caption{Coupling constant $\alpha$ as a function of the sharp
cut-off $k_{max}$ for the $J=0^-$ positronium state with binding
energy $B=0.0225$ a.u.}\label{alpha_kmax_positr}
\end{center}
\end{figure}

This conclusion is confirmed by numerical calculations.
In table~\ref{pos2}
are presented the values of the coupling constant $\alpha$ as a
function of the sharp cut-off $k_{max}$ and for a fixed binding
energy $B=0.0225$. The dependence is very slow -- 0.3\% variation
for $k_{max}\in[10,300]$ -- but it actually corresponds to a
logarithmic divergence of $\alpha(k_{max})$ as it can be seen in
fig.~\ref{alpha_kmax_positr}. 
The origin of this instability is
the coupling to the second component, whose kernel matrix element
$\kappa_{22}$ has an attractive, constant asymptotic limit. If one
removes this component -- which has a very small contribution in
norm -- calculations become stable and give for $\alpha_{NR}=0.30$
the value $\alpha_{LFD}=0.3975$.
\begin{table}[htbp]
\caption{Coupling constant $\alpha$ as a function of the sharp
cut-off $k_{max}$ for the $J=0^-$ positronium state with binding
energy $B=0.0225$ a.u.}\label{pos2}
\begin{tabular}{|c||c|c|c|c|c|c|c|c|c|}\hline
$k_{max}$ &  10  & 20   & 30   &  40  &  50  &  70  & 100  & 200
& 300  \\\hline\hline $\alpha$
&0.3945&0.3928&0.3918&0.3911&0.3905&0.3896&0.3887&0.3867&0.3854\\\hline
\end{tabular}
\end{table}
We should emphasize that as one can see from fig.~\ref{alpha_kmax_positr} and from the table~\ref{pos2},
the development of this instability vs. $k_{max}$ is very slow. The value $k_{max}=300$ (in unites of electron masses)
is very large. At this momentum the contributions having other origin (beyond QED), can make influence and 
change the  behavior of the binding energy vs. $k_{max}$.

Let us now consider also another gauge - the so called light-cone gauge \cite{BL} -- which is often used
in the light-front dynamics calculations. 
In the explicitly covariant version of LFD,  the photon propagator in the light-cone gauge, is obtained from the Feynman one by the replacement (see eq. (2.65) from \cite{cdkm}):
\begin{equation} \label{LCgauge}
-g_{\mu\nu}\to -g_{\mu\nu}+\frac{\omega_{\mu}k_{\nu}+\omega_{\nu}
k_{\mu}}{\omega\cd k}
\end{equation}
The behavior $1/\omega\cd k\sim 1/x$ is singular and should be regularized \cite{BL}.
There are two graphs corresponding to the photon exchange which differ from each other by the order of vertices in the 
light-front time, see e.g. fig.~3 from  \cite{2ferm}. 
The value of the momentum $k$ transferred by photon is different in these LF graphs, see eq. (14) from 
 \cite{2ferm}.  By performing the calculations, we have found that the second term in (\ref{LCgauge}) gives additional contributions to eqs. (\ref{A11pos}) and  (\ref{A22pos}) which turns into:
\begin{eqnarray}\label{A11LC}
A_{11}(\theta,\theta',\gamma)&=&
\int_0^{2\pi}\frac{d\phi}{2\pi D}\left[8\sqrt{\gamma}+\frac{4(1+\gamma^2)\sin\theta\sin\theta'\cos\phi}
{\sqrt{\gamma}|\cos\theta-\cos\theta'|}\right]
\\
A_{22}(\theta,\theta',\gamma)&=&
\int_0^{2\pi}\frac{d\phi}{2\pi D}
\left[  \frac{2}{\sqrt{\gamma}}[(1+\gamma^2)\sin\theta\sin\theta' +2\gamma(1+\cos\theta\cos\theta')\cos\phi]\right.
\nonumber\\
&+&\left.\frac{4(1+\gamma^2)\sin\theta\sin\theta'}
{\sqrt{\gamma}|\cos\theta-\cos\theta'|}\right]
\label{A22LC}
\end{eqnarray}
The singularity $\sim 1/|\cos\theta-\cos\theta'|$ appears from $1/x$ in (\ref{LCgauge}) and, as mentioned, it should be regularized.
In the limit  $\gamma\to 0$, the extra contribution $\sim 1/\sqrt{\gamma}$ in $A_{11}(\theta,\theta',\gamma)$ is smoothen due to the integration over $\phi$, 
whereas it does not change the behavior of $A_{22}(\theta,\theta',\gamma)$ which remains of the form 
$A_{22}(\theta,\theta',\gamma)\propto +1/\sqrt{\gamma}$. As explained above, this corresponds to a spectrum unbounded from below. 
This is manifested  by an unbounded increasing  of the binding energy $B$ as a function of the cutoff $k_{\max}$ 
or by a decreasing -- down to zero -- of the coupling constant $\alpha$ for a fixed value of the binding energy, as it is shown in fig.~\ref{alpha_kmax_positr} and in the table~\ref{pos2}. 
We would like to  again that this dependence on $k_{max}$, fatal for the very existence of stable bound states, 
is very weak and so not at all easy to find its evidence in numerical calculations, specially when using non-uniform mappings.

Due to this very slow $k_{max}$-dependence we have fixed  the cut-off to an arbitrary value $k_{max}=10$
and considered the case $\alpha=3$. 
The non relativistic binding energy is $B=0.0225$ and we found, for the ladder LFD in the Feynman gauge  \cite{2ferm}, a value $B_{LFD} =0.0132$, that is a strong repulsive effect.
This repulsion, observed in most of the kernels examined both for bosons and fermions, 
however contradicts the leading order QED corrections \cite{BS_QM_77} 
\[ B_{QED}=  {\alpha^2\over 4}  \left[1 + \frac{21}{16}\alpha^2 + o(\alpha^4) \right]\approx 0.02516,\]
which are attractive.
This indicates that the ladder light-front kernel, in the Feynman gauge, is unable to predict even the sign for the relativistic corrections of such a genuine
system. It remains to see if this failure is a consequence of  the relative simplicity of the ladder sum or it has other reason.

%%%%%%%%%%%%%%%%%%%%%%%%%%%%%%%%%%%%%%%%%
\section{Conclusions}\label{concl}

In the relativistic framework of Light-Front Dynamics,
we have studied the critical stability of three equal-mass bosons,
interacting via zero-range forces and the two-fermion system interacting via ladder scalar, pseudoscalar and vector  exchanges.

The three equal-mass bosons interact via zero-range forces  constrained
to provide finite two-body mass $M_{2}$. 
We have found that the three-body bound state exists
for two-body mass values in the range $M^{(c)}_2= 1.43\,m \leq M_{2}\leq 2\,m$.
At the zero two-body binding limit, the three-body binding energy is
$B_3^{(c)}\approx 0.012\,m$ and represent the minimal binding energy for a three bosons system with contact interactions.
The Thomas collapse is avoided in the sense that three-body mass $M_3$ is finite, in agreement with \cite{Noyes,tobias}.

However, another kind of catastrophe happens.
Although removing infinite binding
energies, the relativistic dynamics generates zero three-body mass
$M_3$ at a critical value  $M_2=M^{(c)}_2$.
For stronger interaction, i.e. when $0\leq M_{2}< M^{(c)}_2$,
there are no physical solutions of the Light-Front equations with real value of $M_3$.
In this domain, $M_3^2$  becomes negative.

If in the non-relativistic dynamics
the system collapses when its
binding energy tends to $-\infty$, in the relativistic
approach the system does not exist when its mass squared is negative.
This fact can be interpreted as the  relativistic counterpart of the non-relativistic Thomas collapse.

\bigskip
We extended this study to two-fermion system interacting by exchange 
of scalar pseudo scalar and vector particles. In \cite{2ferm}
we have separately examined the different types
of these  couplings and found  very different behaviors
concerning the stability of the  solutions themselves
and their relation with the corresponding non relativistic reductions.

In particular,  the scalar coupling (Yukawa model) is found to be stable  without any kernel
regularization for the $J^{\pi}=0^+$ state and coupling constants below
some critical value $\alpha<\alpha_c=3.72$. For values above $\alpha_c$ the system collapses.
For $J^{\pi}=1^+$ state the solution is unstable.
The comparison with the non relativistic solutions shows always repulsive effects.

Electromagnetic coupling presents the stronger anomalies.
It has been applied to positronium  $0^+$ state. It is found to be unstable
and, once regularized by means of sharp cut-off, the ladder approximation in the Feynman gauge gives relativistic
corrections of opposite sign compared to QED perturbative results.
This failure shows, probably, the poorness of the ladder approximation
in one of the rare cases in which it can be confronted to experimental results.
\bigskip

As a final remarks, we would like to emphasize again that, as it  was first pointed out  in \cite{Noyes}, the relativistic dynamics allows to exist, in principle, 
systems which would not exist in the non-relativistic framework. 
Their existence is determined by the properties and strength of relativistic interaction and we dentoe this fact by "critical stability". 
The pioneering work of Pierre Noyes \cite{Noyes} opened thus a fruitful and interesting field  
in the theory of few-body systems.

%%%%%%%%%%%%%%%%%%%%%%%%%%%%%%%%%%%%%%%%%

\end{document}